# Electric-field sensing with driven-dissipative time crystals in room-temperature Rydberg vapor

Darmindra Arumugam

Jet Propulsion Laboratory, California Institute of Technology, Pasadena, 91109, California, USA

E-mail: darmindra.d.arumugam@jpl.nasa.gov;

**Abstract**

Mode competition in nonequilibrium[1-2] Rydberg gases enables the exploration of emergent many-body phases[3-5]. This work leverages this emergent phase for electric field detection at room temperature. Sensitive frequency-resolved electric field measurements at very low-frequencies (VLF) are of central importance in a wide range of applications where deep-penetration is required in communications[6,7], navigation[8] and imaging or surveying[9,10]. The long wavelengths on order of 10-100 km (3-30 kHz) limit the efficiency, sensitivity, and bandwidth of compact classical detectors that are constrained by Chu's limit[11]. Rydberg-atom electrometers[12-14] are an attractive approach for microwave electric-field sensors but have reduced sensitivity at lower-frequencies. Very recent efforts to advance the standard Rydberg-atoms approach is based on DC electric-field (E-field) Stark shifting[15,16] and have resulted in sensitivities between 67.9-2.2 uVcm$^{-1}$Hz$^{-1/2}$ (0.1-10 kHz) by fine optimization of the DC E-field. A major challenge in these approaches is the need for embedded electrodes or plates due to DC E-field Stark screening effect[17], which can perturb coupling of VLF signals when injected from external sources. In this article, it is demonstrated that state-of-art sensitivity (~1.6-2.3 uVcm$^{-1}$Hz$^{-1/2}$) can instead be achieved using limit-cycle oscillations in driven-dissipative Rydberg atoms by using a magnetic field (B-field) to develop mode-competition between nearby Rydberg states. The mode-competition between nearby Rydberg-states develop an effective transition centered at the oscillation frequency capable of supporting external VLF E-field coupling in the ~10-15kHz regime without the requirement for fine optimization of the B-field magnitude.

The interplay of driving and dissipation in quantum nonequilibrium systems[1] gives rise to emergent many body phases, central to atomic physics[2] and pivotal for novel detection paradigms. Recent evidence[3-5] shows that mode-competition in nonequilibrium Rydberg gases provide a means to study many-body phases. This emergent phase is exploited in this article for sensitive electric field detection at room-temperature.

High-resolution electric field (E-field) measurements at very low frequencies (VLF) are crucial for numerous applications that require deep signal penetration, including in communications[6,7], navigation[8], and geophysical imaging or subsurface surveying[9,10]. Many applications require both sensitive field detection and frequency-resolved measurements. The very long wavelengths, ranging from 10-100 km for VLF signals at 3–30 kHz, impose constraints on the efficiency, sensitivity, and bandwidth when portable/compact electrically small ($\ll \lambda$) classical detectors or antennas are used due to Chu's limit[11].

Atomic sensors leverage highly coherent quantum systems to probe atoms and detect weak signals with exceptional sensitivity and precision[18]. Alkali atoms, such as cesium (Cs) and rubidium (Rb), which exhibit high vapor pressure, can be excited to Rydberg states with high principal quantum numbers, making them sensitive to microwave and millimeter-wave fields[12-14]. Typically, two or more lasers are employed to prepare and probe these atoms. This process involves creating transparency in the probe laser through electromagnetically induced transparency (EIT)[19] and monitoring perturbations in the EIT spectrum caused by external microwave fields. On-resonance transitions for VLF E-field detection at kHz frequencies are impractical with Rydberg atoms because the energy associated with kHz-frequency photons is far too small to directly drive transitions between states, necessitating alternative techniques such as off-resonant or indirect methods.

Recent advances in VLF E-field detection with Rydberg atoms have achieved high-resolution measurements using off-resonant techniques with DC electric-field (E-field) Stark shifting, demonstrating sensitivities ranging from 67.9-2.2 uVcm⁻¹Hz⁻¹/² (0.1-10 kHz)[15,16]. The applied DC E-field enhances sensitivity by inducing a Stark shift, which increases the polarizability of the selected Rydberg state, enabling detection of RF/AC fields through modulation of the Rydberg-EIT spectrum. A significant challenge in these methods is the requirement for embedded electrodes or plates inside the vapor cell to overcome the DC E-field Stark-screening effect[17]. The embedded electrodes or plates interfere with or negate the coupling of very low-frequency (VLF) signals when introduced from external sources. In addition, the DC E-field must be carefully optimized[15,16] to induce a Stark shift that enhances the polarizability and sensitivity, while avoiding excessive perturbations that could reduce measurement accuracy or interfere with the detection of weak RF/AC fields.

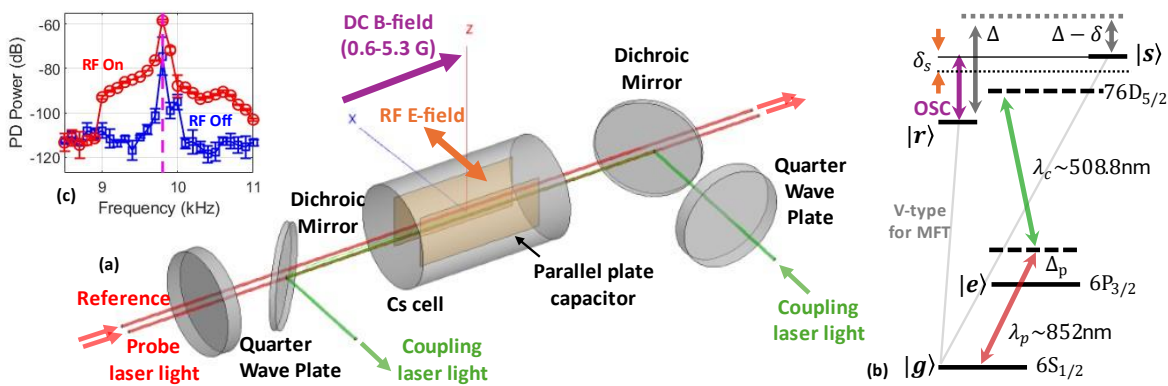

Fig. 1: (a) Experimental protocol, energy diagram, and swept RF frequency spectrum. (a) The probe and coupler laser light are counter-propagated in a Cesium vapor cell to excite atoms to the Rydberg state. A background DC magnetic field between 0.6-5.3 G is generated using a Helmholtz coil (not shown here). A parallel plate capacitor is embedded in the vapor cell, which is used for uniform RF field generation inside the vapor cell. Quarter wave plates are used to fine tune the polarization state of the probe and coupler laser light. A beam displacer is used to develop a reference probe beam for balanced detection to reduce technical noise. (b) The probe drives atoms to the first excited state at $6P_{3/2}$ and is detuned by $\Delta_p/2\pi$ = 22MHz. The coupler drives the atoms to a principal quantum number of n = 76 at $76D_{5/2}$ and is detuned by $\Delta_c/2\pi$ = 30MHz. When the magnetic field is tuned between approximately 4-5G, mode competition is observed between nearby sublevels and give rise to limit cycle oscillatory dynamics (OSC). A V-type model (grey) is used for mean-field treatment (MFT) in Methods. (c) The oscillatory dynamics couple the states in a synchronous manner that is observed as a resonance in the probe measured signal at about ~9.8kHz. An injected weak RF signal via the parallel plate capacitor that is close to this resonance [see $\delta_s$ in (b)] modulates the probe signal as the system is sensitized to nearby frequencies. A swept RF frequency shows a finite RF detection bandwidth with a highest sensitivity close to the OSC resonance frequency.

Recent research has shown that in a driven-dissipative Rydberg vapor, synchronization of atomic oscillations can emerge due to strong global coupling via the mean Rydberg density, which causes frequency and phase locking among velocity classes despite atomic motion[3-5]. This results in self-sustained limit cycle oscillations (OSC) through a Hopf bifurcation, characterized by mode competition between nearby states and periodic oscillations in bulk vapor properties. The observed oscillation frequencies (via probe laser transmission), typically in the range of 10–25 kHz, are robust over long timescales and depend on system parameters, including Rabi frequencies, field intensities, vapor density, and interaction strengths. The observed OSC and persistent periodic behavior can be interpreted as a manifestation of a dissipative time crystal[3], where spontaneous time-translation symmetry breaking arises via competition between Rydberg states and is stabilized by dissipation. Further, it has been shown that this time crystal state can exist at room temperatures[3].

In this article, it is demonstrated that state-of-art sensitivities between 1.6-2.3 uVcm⁻¹Hz⁻¹/² can instead be achieved using OSC in driven-dissipative Rydberg atoms at room temperature by using a magnetic field (B-field) to develop mode-competition between nearby Rydberg states[3]. The mode-competition between nearby Rydberg-states develop an effective transition centered at the OSC frequency capable of supporting external VLF E-field coupling at ~10-15kHz without the requirement for fine optimization of the B-field magnitude or use of DC E-fields for Stark shifting. While the present work addresses VLF detection, the general technique could be useful in super heterodyning for microwave detection and will be investigated in the future.

## Results

### Observed OSC and RF field driven perturbations

A probe (~852nm) and coupler (~509nm) laser that is counter-propagated is used to interact with alkali atoms

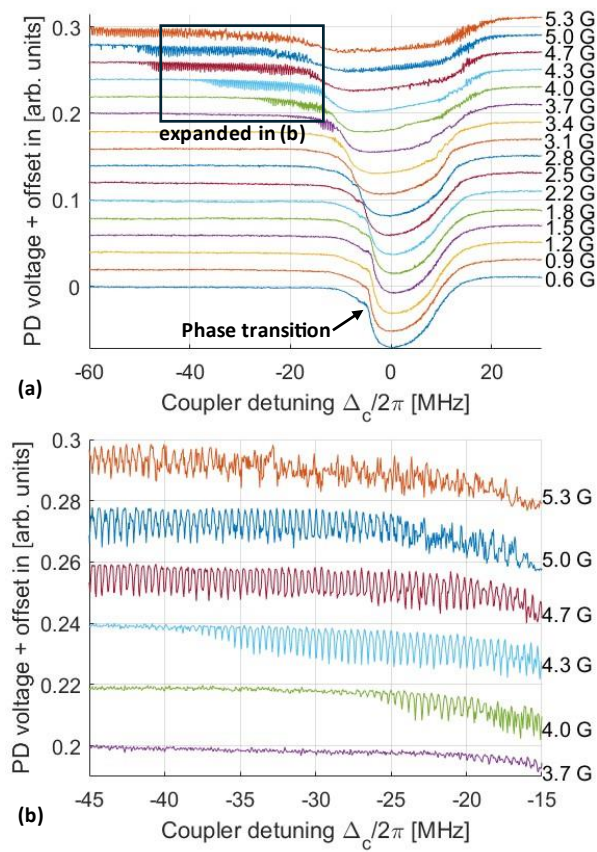

**(a)**

**(b)**

**Fig. 2:** (a) EIT spectrum (arbitrary units) as a function of coupler detuning and magnetic flux density with room-temperature Cesium Rydberg vapor. The scanning rate is 2π × 4.88 MHz/ms. Rabi frequencies of the coupler and probe were held constant. Details of setup is given in Fig. 1 and described in Methods. (b) An expanded view of the black box region highlighted in (a) shows that the magnetic flux density magnitude impacts oscillatory magnitude observed by the probe transmission, and that the stability range of the limit cycle oscillations (OSC) in the coupler detuning space.

(Cs-133) in a vapor cell at room temperature as shown in Fig. 1a. Both probe and coupler light pass through a quarter-wave plate (QWP) that is used to generate and tune elliptical polarization states of the light. A DC B-field between 0.6-5.3 G is generated parallel in direction to the interaction axis. The vapor cell has an embedded parallel-plate capacitor (PPC) that simplifies RF E-field injection. Here a voltage and a transfer function are used to develop an RF E-field in the PPC. Fig. 1b shows the energy diagram for the atomic system. The probe drives atoms to the first excited state at $6P_{3/2}$ and is detuned by $\Delta_p/2\pi = 22$MHz. The coupler drives the atoms to a principal quantum number of n = 76 at $76D_{5/2}$ and is detuned by $\Delta_c/2\pi = 30$MHz. At this sufficiently high[3] n, a phase transition emerges (Fig. 2a) that results in stable oscillations (OSC) due to mode competition when a magnetic field between ~4.3-4.7 G is introduced.

The OSC is considered a coupling between $|r\rangle$ and $|s\rangle$ (states that are competing, see Fig. 1b), and this develops a competition dependent RF transition that can be used to sense RF fields – The self-sustained oscillations represent a resonant mode of the atomic system. If an external RF field with a frequency near the OSC frequency is applied, it can couple to the system's oscillatory mode, resulting in a modulation or enhancement of the system's response. While the OSC frequency does not correspond directly to a traditional atomic transition, it arises from the collective dynamics of the system. These collective oscillations can interact with an external RF signal at or near the OSC frequency, manifesting as a resonant response in the system. Fig. 1c shows the RF spectrum of the probe transmission signal (both lasers are frequency locked). The OSC spectrum (blue) is centered at $f_{OSC}$~9.8kHz. An injected RF signal via PPC near $f_{OSC}$ modulates the probe transmission, which is sensitive to nearby frequencies. Sweeping the RF frequency reveals a finite bandwidth, with the highest signal-to-noise (SNR) occurring close to the $f_{OSC}$ resonance. Details of the: 1) Experimental setup is given in Methods: Experimental setup and approach, 2) System architecture and design is given in Supplementary Section 1, 3) Vapor cell design and magnetic field perturbation analysis in Supplementary Section 2, Impedance and transfer function for the PPC is in Supplementary Section 3.

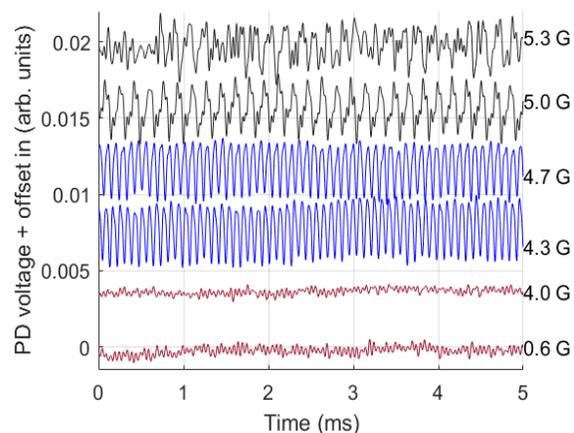

**Fig. 3:** Probe transmission as a function of time for different magnetic flux density magnitudes. The probe, coupler Rabi frequency was set to $\Omega_{p,c}/2\pi = 13.97$ MHz, 0.85 MHz, the laser wavelengths to drive $6S_{1/2}-6P_{3/2}-76D_{5/2}$, with $\Delta_{p,c}/2\pi = 22$ MHz, 30 MHz (frequencies locked to observe transients). Between B = 4.3-4.7G (blue lines), the limit cycle observations (OSC) are stable and sustained. Below B = 4G (red lines), OSC disappears, and background probe laser noise dominates. Above B = 5G (black lines), the OSC is unstable.

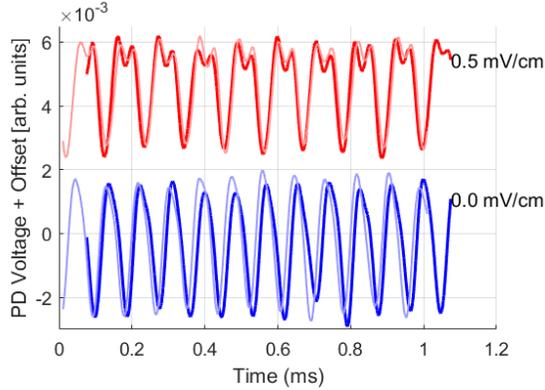

Fig. 4: The RF modulation of limit cycle oscillations (OSC) as observed by probe transmission signal. Blue curves are OSC probe transmission with no RF signal, and red curves are a modulated OSC probe transmission due to RF modulation via an electric field. The electric field (red curves) at f = 9.3 kHz and 0.5mV/cm was injected using a parallel-plate capacitor (see Fig. 1 and Methods). Data collection was repeated to show consistency (lighter shaded colors).

To observe OSC, we investigate the EIT spectrum at high-n with added B-field. Fig. 2a gives the EIT spectrum (of the probe transmission) as a function of coupler detuning. A phase transition that results in a limit cycle oscillation (black box). Coupler detuning sweep direction is positive (increasing detuning frequency), and the scanning rate is $2\pi \times 4.88$ MHz/ms. Coupler and probe Rabi frequencies was $\Omega_c = 0.85$ MHz and about $\Omega_p = 13.97$ MHz. Fig. 2b gives an expanded view of the black box region in Fig. 2a. Stable periodic oscillations are observed over large coupler detuning ranges within B-field magnitudes of ~4.3-4.7 G.

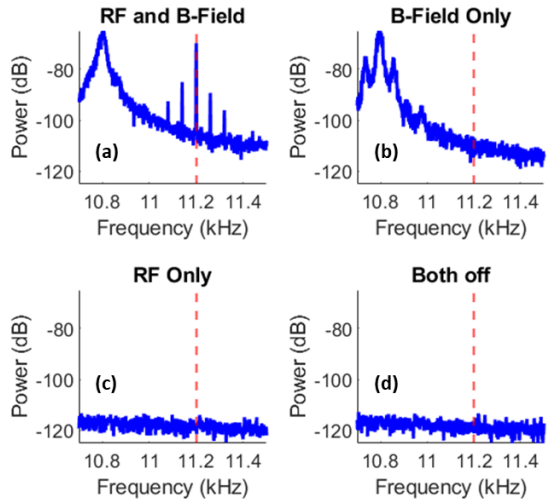

Fig. 5: Spectral analysis of a driven dissipative time crystal with a strong RF field. (a) Both fields (RF and magnetic) are turned on, the limit cycle oscillations are observed, and the strong RF at 11.2kHz is observed. RF and B field magnitudes are ~0.3mV/cm and ~4.5G. (b) The RF is turned off and only B-field is on. Here the RF signal is no longer observed in the spectrum. (c) The RF is turned on and the B field is off. (d) Both fields are turned off. In (c) and (d), the measured power throughout the band shown is limited by probe laser noise.

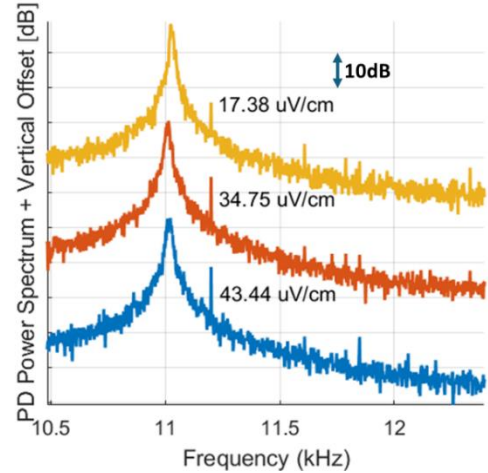

Fig. 6: An example of progressively weaker RF fields at 11.2kHz. The probe Rabi frequency and magnetic field magnitude is tuned very finely to increase the quality factor of the limit cycle oscillation (OSC). Data collected with a 10Hz resolution bandwidth. Three example curves shown to exemplify the dependence on RF field strength. Curves are shifted vertically (vertical offset) for display.

The periodicity is best observed when lasers are frequency locked and, in the time-domain. Fig. 3 shows the probe transmission as a function of time for varying B-field magnitudes. The lasers are frequency locked, and B-field turned on to observe OSC and periodicity of the mode competition. Time axis referencing is started after onset of OSC, and thus early transients are not captured. Between ~4.3-4.7G (blue lines), a stable and sustained OSC is observed. Below ~4G (red lines), the OSC vanishes, and background probe laser noise dominates. Above ~5G (black lines), the OSC becomes unstable and is not periodic. The $f_{OSC}$ was found to be between 9-12kHz and sensitive to Rabi frequencies of the probe and coupler. To exemplify the modulation of OSC via RF signal injection, the B-field magnitude is set to ~4.5G and find $f_{OSC}$~11.72kHz (Fig. 4) as observed by the probe transmission signal. The blue curves in Fig. 4 represent the OSC probe transmission without an RF signal, while the red curves show the modulated OSC probe transmission resulting from a strong RF-field modulation via an applied electric field. The electric field, at $f_{RF}$=9.3kHz and at ~0.5mV/cm was introduced using a parallel-plate capacitor for convenience (see Fig. 1 and Methods). The resulting time-domain probe transmission (red curve) is perturbed and shows a dominant 9.3kHz signal. Light colored curves are repeated measurements. This is further illustrated in a separate experiment where a strong RF-field and B-field is sequentially turned on/off as shown in Fig. 5. Here the probe Rabi and B-field is tuned to generate an OSC at ~10.8kHz (Fig. 5b, B-field only). When a strong RF is switched on at 11.2kHz, a tone with some higher order perturbations is observed centered at ~11.2kHz due to the injected RF-field (Fig. 5a). When the B-field is turned off (Fig. 5c,d), background probe transmission noise due to probe laser noise is measured across the entire band, regardless of whether RF-field is on/off (Fig. 5c/d).

It is found generally that slight tuning of the Rabi frequencies and B-field magnitude can alter the quality-factor (Q-factor) of the OSC. As the Rabi frequencies are slightly tuned, they shift the resonance conditions for the mode competition. This can alter the balance between energy transfer and dissipation, thereby affecting both the oscillation frequency and the damping (related to the Q factor). The degree of coupling between the modes depends on the spacing of the energy levels. Slight tuning of the B-field can enhance or suppress this coupling, affecting the persistence and sharpness of the oscillations. Fig. 6 shows a tuned Q-factor at $f_{OSC}$~11.02kHz (compare to Fig. 5b). In Fig. 6, a weak RF-field is additionally injected (17.3,34.7,43.4uV/cm) at $f_{RF}$~11.2kHz. The observed spectrum shows both the OSC and RF signals, with progressively lower SNR for weaker RF signals.

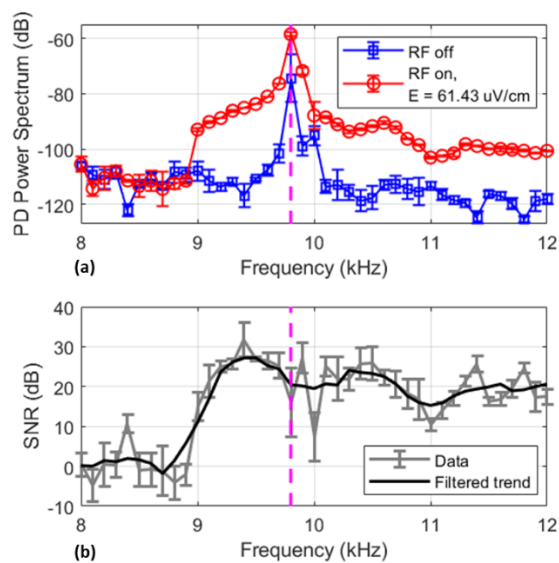

Fig. 7: Swept frequency analysis. (a) The electric field transmit frequency is varied over the range of 8-12kHz with a field intensity of ~61.43uV/cm (red curve). This is compared to the case when the field is turned off (blue curve). A passband is observed that drops off rapidly at lower frequencies f<$f_{OSC}$, but slowly as f>$f_{OSC}$. (b) Interpreted signal-to-noise (SNR) ratio showing a peak SNR ~ f<$f_{OSC}$ ($f_{OSC}$~9.8kHz and $f_{peak}$~9.5kHz).

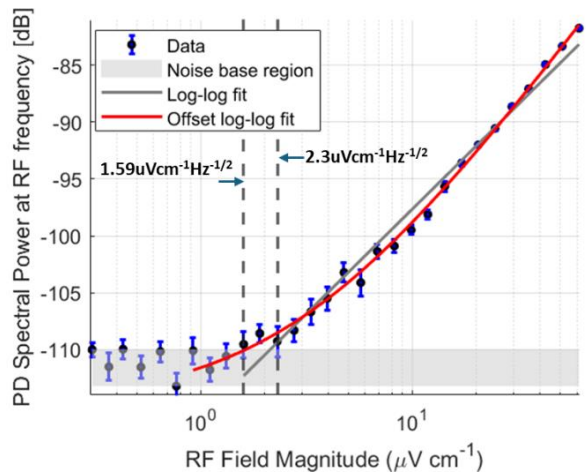

Fig. 8: Measured probe signal transmission power at the RF frequency for variation in RF field magnitudes. The noise base region (gray rectangular box) is estimated, and two best fit lines are included. The sensitivity is estimated based on data points that are closest but above the noise base region highlighted and based on a 1Hz resolution bandwidth used in the measurement. Two points are selected based on proximity to the best fit lines. These are found to be 1.59uVcm⁻¹Hz⁻¹/² and 2.3uVcm⁻¹Hz⁻¹/².

### Detection bandwidth of effective transition and sensitivity to injected RF E-field

The atomic system becomes sensitive to external RF signals near $f_{OSC}$ due to the resonant nature of the limit cycle oscillations. This sensitivity manifests as a modulation (strong RF) or perturbation (weak RF) of the system's oscillatory dynamics. While this is not a traditional atomic transition, an effective collective resonance/dynamical transition is developed. From the perspective of sensitive E-field detection, it is useful to study the detection bandwidth of this transition and the regime of highest sensitivity for the setup presently used. Fig. 7 shows a swept injected RF frequency analysis. The RF frequency is varied from 8 to 12 kHz with a field intensity of ~61.4 μV/cm (red curve) and compared to the case where the field is turned off (blue curve) in Fig. 7a. The magenta dashed lines correspond to $f_{OSC}$~9.8kHz. A passband is observed, with a sharp drop at frequencies f<$f_{OSC}$ and a more gradual decline for f>$f_{OSC}$. The estimated SNR as a function of frequency is given in Fig. 7b. Peak SNR was found at $f_{peak}$~9.5kHz and the -10dB bandwidth was found to be ~1.7kHz based on a moving average filter to obtain a smooth trend (black curve in Fig. 7b). To study sensitivity of the detector close to OSC, the RF E-field magnitude is swept from ~0.3-61.4 uV/cm for $f_{RF}$=9.5kHz. RF voltage applied is converted to field magnitude based on PPC geometry and a transfer function (see Supplementary Section 3).

The peak power measured at $f_{RF}$ in the power spectrum of the probe transmission signal is shown in Fig. 8 in log-log scale. A resolution bandwidth of 1Hz is used. The gray-region is the estimated noise-base, dominated by probe laser noise. Measurements are repeated 4 times over two days, and the error bars give the standard error of the mean. Two best curve fits to the data points are presented (gray line for log-log fit, and a red-curve for an offset log-log fit). The red/gray fitted-curves 4 times give residual RMS (root mean square) errors of 2.03/0.63dB respectively. The sensitivity is estimated between 1.59-2.3uVcm$^{-1}$Hz$^{-1/2}$, based on the mean data points closest to the red-gray curves and above the system noise-base. Details including curve-fit methods and the resulting estimated errors are given in Supplementary Section 4.

## Discussion

In Rydberg gases, collective oscillations or limit cycles arise from the interplay of many interacting atoms and are not directly tied to the single-particle energy levels of individual atoms. The oscillations act as an effective collective resonance/dynamical transition, driven by nonlinear interactions and mode competition. At the self-oscillation frequency (OSC), the atomic system becomes sensitive to external RF signals near the OSC frequency due to the resonant nature of the system. This RF sensitivity manifests as a modulation or perturbation of the system's oscillatory dynamics, even though it may not correspond to a direct atomic transition. In this article, it is demonstrated that this effective transition can be used as a sensitive RF E-field sensor in the VLF frequency range. It is shown that a state-of-art sensitivity between 1.59-2.3uVcm$^{-1}$Hz$^{-1/2}$ at ~9.5kHz can be obtained when a B-field of about 4.5G is used to drive mode competition and oscillations. The detection bandwidth appears to be narrow (~1.7kHz for -10dB), so tuning of the OSC frequency may be needed for detection at different VLF frequencies and will be studied as future work.

The technique does not require careful optimization of the B-field, and as a result is simpler to use when compared to DC E-field Stark-shift based techniques in recent articles[15,16] that require optimization. In addition, a notable challenge in DC E-field Stark-shift methods is the need for embedded electrodes/plates within the vapor cell to mitigate Stark-screening effects[15,16], which have a cutoff frequency as high as a few kHz even when suitable cells such as monocrystalline sapphire vapor cells are used[17]. These electrodes/plates can disrupt or negate the coupling of VLF signals introduced from external sources. While we use an embedded PPC for convenience, it is not needed as a DC E-field is not used in the present technique.

## Methods

### Mean-field dynamics with RF near OSC frequencies

The simplified model developed is aimed at providing a qualitative description of RF coupling with quantitative predictions of RF bandwidth of the emergent transition. The qualitative approach begins with a characteristic Hamiltonian for a V-type three-level system[3] (see Fig. 1b) modified to include coupling between the excited states $|r\rangle$ and $|s\rangle$ via an external RF field. The field is slightly detuned relative to the energy between excited states. The Hamiltonian in the rotating frame is:

$$\hat{H} = \frac{\Omega}{2}\sum_i \left(\hat{\sigma}_{gr}^i + \hat{\sigma}_{gs}^i + \text{H.c.}\right) - \sum_i (\Delta_r \hat{n}_r^i + \Delta_s \hat{n}_s^i)$$
$$+ \frac{1}{2}\sum_{i \neq j} V_{ij}\left(\hat{n}_r^i \hat{n}_r^j + 2\hat{n}_r^i \hat{n}_s^i + \hat{n}_s^i \hat{n}_s^j\right)$$
$$+ \frac{\Omega_{rs}}{2}\sum_i \left(\hat{\sigma}_{rs}^i e^{i\delta_s t} + \hat{\sigma}_{sr}^i e^{-i\delta_s t}\right),$$

where $\hat{\sigma}_{\alpha\beta}^i = |\alpha^i\rangle\langle\beta^i|$ is the transition operator ($\alpha, \beta = g, r, s$), and $\hat{n}_\alpha^i = |\alpha^i\rangle\langle\alpha^i|$ ($\alpha = r, s$) denotes the local Rydberg density. $\Omega$ is the ground-to-excited state Rabi frequency and $\Omega_{rs}$ the direct Rabi coupling between the excited states $|r\rangle$ and $|s\rangle$. $\delta_s$ is the detuning frequency of the drive RF field. Equal interaction strengths, $V_{ij}$, is assumed to simplify development. The evolution of an operator $\hat{O}$ is governed by $\partial_t\langle\hat{O}\rangle = i\langle[\hat{H}, \hat{O}]\rangle + \mathcal{L}^a[\hat{O}]$, where the Lindblad term accounts for the decay of the Rydberg states: $\mathcal{L}^a[\hat{O}] = \sum_i \gamma\left(2\,\hat{\sigma}_{\alpha g}^i \hat{O}\hat{\sigma}_{g\alpha}^i - \{\hat{n}_\alpha^i, \hat{O}\}\right)$ ($\alpha = r, s$). Considering the added coupling between the excited states $|r\rangle$ and $|s\rangle$, the equations of motion for the first moments are given by the population evolution equations describing the balance between excitation, decay, and excited-state coupling:

$$\frac{d}{dt}\langle\hat{n}_r^i\rangle = i\frac{\Omega}{2}\left(\langle\hat{\sigma}_{gr}^i\rangle - \langle\hat{\sigma}_{rg}^i\rangle\right) - \gamma\langle\hat{n}_r^i\rangle$$
$$+ i\frac{\Omega_{rs}}{2}\left(\langle\hat{\sigma}_{rs}^i e^{i\delta_s t}\rangle - \langle\hat{\sigma}_{sr}^i e^{-i\delta_s t}\rangle\right)$$
$$\frac{d}{dt}\langle\hat{n}_s^i\rangle = i\frac{\Omega}{2}\left(\langle\hat{\sigma}_{gs}^i\rangle - \langle\hat{\sigma}_{sg}^i\rangle\right) - \gamma\langle\hat{n}_s^i\rangle$$
$$- i\frac{\Omega_{rs}}{2}\left(\langle\hat{\sigma}_{rs}^i e^{i\delta_s t}\rangle - \langle\hat{\sigma}_{sr}^i e^{-i\delta_s t}\rangle\right),$$

and by the coherence evolution equations, influenced by interactions and Rabi oscillations, both with additional terms due to effects of $\Omega_{rs}$ (from the RF field):

© 2025. California Institute of Technology. Government sponsorship acknowledged.

$$\frac{d}{dt}\langle\hat{\sigma}_{gr}^i\rangle = i\frac{\Omega}{2}\big(2\langle\hat{n}_r^i\rangle + \langle\hat{n}_s^i\rangle + \langle\hat{\sigma}_{sr}^i\rangle - 1\big)$$
$$+ i\left(\Delta_r - \sum_{j\neq i} V_{ij}(\langle\hat{n}_r^j\rangle + \langle\hat{n}_s^j\rangle) + i\frac{\gamma}{2}\right)\langle\hat{\sigma}_{gr}^i\rangle$$

$$\frac{d}{dt}\langle\hat{\sigma}_{gs}^i\rangle = i\frac{\Omega}{2}\big(2\langle\hat{n}_s^i\rangle + \langle\hat{n}_r^i\rangle + \langle\hat{\sigma}_{rs}^i\rangle - 1\big)$$
$$+ i\left(\Delta_s - \sum_{j\neq i} V_{ij}(\langle\hat{n}_r^j\rangle + \langle\hat{n}_s^j\rangle) + i\frac{\gamma}{2}\right)\langle\hat{\sigma}_{gs}^i\rangle$$

$$\frac{d}{dt}\langle\hat{\sigma}_{rs}^i\rangle = i\frac{\Omega}{2}\big(\langle\hat{\sigma}_{gs}^i\rangle - \langle\hat{\sigma}_{rg}^i\rangle\big) - i(\Delta_r - \Delta_s - i\gamma)\langle\hat{\sigma}_{rs}^i\rangle$$
$$+ i\frac{\Omega_{rs}}{2}\big(\langle\hat{n}_r^i\rangle - \langle\hat{n}_s^i\rangle\big)e^{i\delta_s t}.$$

In the mean-field treatment[20-22], it is assumed that the correlation between different atoms is negligible and higher-order moments are factorized as $\langle\hat{n}_r^i\hat{\sigma}_{gr}^j\rangle \approx \langle\hat{n}_r^i\rangle\langle\hat{\sigma}_{gr}^j\rangle$, $\langle\hat{n}_s^i\hat{\sigma}_{gs}^j\rangle \approx \langle\hat{n}_s^i\rangle\langle\hat{\sigma}_{gs}^j\rangle$. Assuming a uniform spatial distribution, where $\langle\hat{n}_r^i\rangle = n_r$ and $\langle\hat{n}_s^i\rangle = n_s$, and defining $E_{NL} = \chi(n_r + n_s)$, which represents the nonlinear energy shift due to atomic interactions. The mean-field population and coherence equations are:

$$\dot{n}_r = \frac{\Omega}{2}\Im(\sigma_{gr}) - \gamma n_r + \frac{\Omega_{rs}}{2}\Im(\sigma_{rs}e^{i\delta_s t})$$
$$\dot{n}_s = \frac{\Omega}{2}\Im(\sigma_{gs}) - \gamma n_s - \frac{\Omega_{rs}}{2}\Im(\sigma_{rs}e^{i\delta_s t})$$
$$\dot{\sigma}_{gr} = i\frac{\Omega}{2}(2n_r + n_s + \sigma_{sr} - 1)$$
$$+ i\left(\Delta_r - E_{NL} + i\frac{\gamma}{2}\right)\sigma_{gr}$$
$$\dot{\sigma}_{gs} = i\frac{\Omega}{2}(2n_s + n_r + \sigma_{rs} - 1)$$
$$+ i\left(\Delta_s - E_{NL} + i\frac{\gamma}{2}\right)\sigma_{gs}$$
$$\dot{\sigma}_{rs} = i\frac{\Omega}{2}(\sigma_{gs} - \sigma_{rg}) - i(\Delta_r - \Delta_s - i\gamma)\sigma_{rs}$$
$$+ i\frac{\Omega_{rs}}{2}(n_r - n_s)e^{i\delta_s t}.$$

The inclusion of $\Omega_{rs}$ (via external RF) modifies both the population and coherence dynamics as follows: (1) The terms $\pm(\Omega_{rs}/2)\Im(\sigma_{rs}e^{i\delta_s t})$ introduce coherent population transfer between $n_r$ and $n_s$, driven by the RF field; (2) The coherence term $\sigma_{rs}$ now evolves with an explicit driving term $i(\Omega_{rs}/2)(n_r - n_s)e^{i\delta_s t}$, linking it directly to the imbalance between the two excited states. This introduces RF-driven coherence oscillations that modulate the interference effects observed in the system; (3) Since $E_{NL}$ depends on $n_r + n_s$, the self-consistent feedback from the RF-driven population dynamics can shift the effective detunings $\Delta_r$ and $\Delta_s$. This means that the system's natural oscillation frequencies and potential limit-cycle behaviors may also potentially be tuned dynamically by $\Omega_{rs}$ and $\delta_s$.

The focus here is to study temporal and spectral nature of the solutions and to inspect RF sensing bandwidth of the transition created by the OSC. Using a numerical solution, the time evolution of the system of equations is computed, and $\Im(\sigma_{gr})$ is extracted, which represents the response of the probe transmission from the ground-to-intermediate state in the experiments. Fig. 9a shows three cases for $\Im(\sigma_{gr})$ without OSC (red), with OSC (blue), and with OSC and added strong RF (green) to perturb the OSC. Decay rate used is estimated at $\gamma/2\pi$=31.66 kHz consisting of contributions from spontaneous decay and transient time broadening. Due to two-photon scheme, $\Omega = \Omega_c\Omega_p(1/\Delta_p - 1/\Delta_c)/2$ is estimated at 0.072 MHz (see Methods: Experimental setup). $\delta/2\pi$=0.32MHz and $\chi/2\pi \sim 1.2$MHz was tuned to get OSC frequency close to 9-11kHz. $\Delta/\gamma \sim \pm 5$, and $\Omega/\gamma \sim 2.3$ is estimated from the experiments and calculations. Based on energy splitting calculations we expect $\delta/2\pi$ closer to ~1MHz and this discrepancy will be studied in the future. $\Delta = \pm \delta/2$ is used, and sign + is expected to show no OSC (red in Fig. 9a). When outside the OSC regime, $\Im(\sigma_{gr})$ decays rapidly, whereas inside the OSC regime a sustained oscillation emerges, and it can be perturbed by an RF field close to the OSC frequency. For weak RF signals which are detuned in frequency relative to OSC, spectral decomposition shows strong peaks at the detuning frequency corresponding to the RF signal. As higher RF frequency detuning ($\delta_s$, see Fig. 9b) is used, a rapid drop-off in RF signal power is seen in the simulated data (Fig. 9c). Qualitatively, there is a good agreement with the experimental results in Fig. 7, indicating a presence of a transition near the OSC that supports efficient RF signal coupling. However, quantitative estimation based on the current theory for RF detection bandwidth based on Fig. 9c is between ~0.3-0.74 kHz (-10dB full-width bandwidth from first or second data point), compared to about ~1.7kHz from experimental data in Fig. 7.

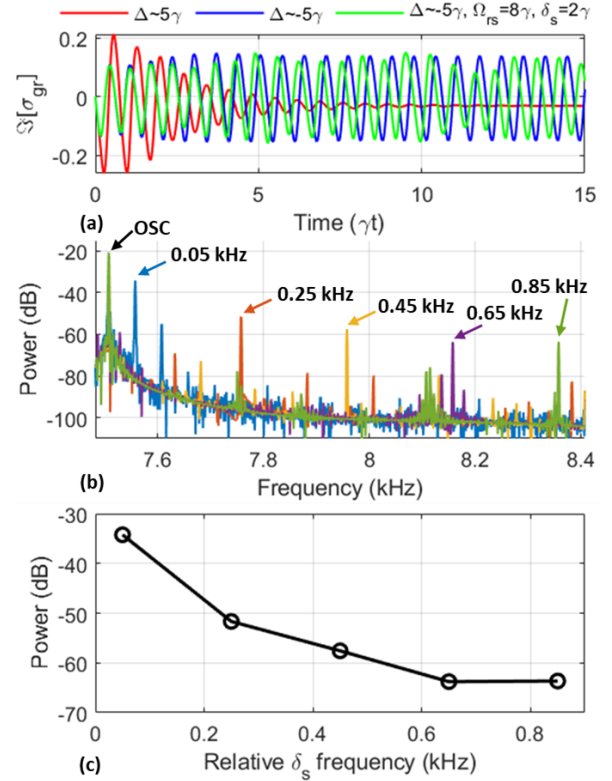

Fig. 9: Theoretical simulations of the imaginary part of $\sigma_{gr}$ based on the mean-field treatment. (a) Time domain response of the system without OSC (red) showing rapid decay, with OSC (blue) showing sustained oscillations, and OSC that is perturbed by a strong RF field ($\Omega_{rs} = 8\gamma$) close to OSC frequencies (green). (b) The spectrum of multiple simulations using a weak RF field ($\Omega_{rs} = 0.04\gamma$) with different RF detuning $\delta_s$. (c) Drop-off in signal power as a function of detuning from (b) showing a rapid attenuation and indicating a presence of transition supporting RF coupling.

**Experimental setup and approach**

Multiple experiments were conducted to study the EIT spectrum, magnetic flux density dependence, RF field magnitude and frequency dependence, sensitivity (Figs. 2-8), and for a characteristic spectral measurement for comparison to a simplified simulation (Fig. 9). The setup for the Rydberg atomic system was identical in all cases (see Fig. 1 and Supplementary Section 1).

The vapor cell used had a total length of 56.6mm and diameter of 35mm, with embedded parallel plates that are 18mm x 45mm each, and the distance between plates are 15mm (see Supplementary Section 2 and Extended Data Figure 2). The energy level configuration (Fig. 1b) is realized by $6S_{1/2}$-$6P_{3/2}$-$76D_{5/2}$, and a B-field magnitude between 4.3-4.7 G to drive the limit cycle oscillations (OSC). The B-field was generated by a 30cm diameter Helmholtz coil, spaced 15cm apart, by applying a DC voltage between 1.2-1.5V. This B-field penetrated the vapor cell with embedded parallel-plate capacitors (PPC). The B-field was measured using a secondary magnetometer to confirm accuracy of magnitudes noted in the study. Method of Moments electro-magnetic simulations were used to study perturbation of the B-field due the PPC, and it was shown (see Supplementary Section 2) that the B-field had only a perturbation of ~0.4% in the presence of the PPC. To study injected RF signal dependence (magnitude and frequency), the PPC was used to drive RF fields into the vapor cell (see Fig. 1) using a low-noise RF signal generator. Impedance transfer functions were measured with a RF impedance analyzer and a fitted circuit model (see Supplementary Section 3) was employed to determine the voltage across the PPC and corresponding E-field magnitude generated by the PPC.

The two lowest states were $6S_{1/2}$ F=4 and $6P_{3/2}$ F' = 4, driven by the probe laser. The frequency of the probe laser was locked to the hyperfine structure via a standard saturation spectroscopy technique, giving a probe $\lambda_p \sim 852.35672$nm. Probe linewidth is estimated to be <58kHz, and it's detuning is $\Delta_p/2\pi$ = 22MHz. The probe Rabi frequency and beam diameter was about $\Omega_p$ = 13.97 MHz and 1.1 mm ($1/e^2$ diameter). A half-wave plate is used to set the polarization state of the light prior to a Calcite beam displacer. This state was maintained slightly elliptically polarized using a tunable quarter-wave plate ($\lambda/4$), which is optimized to maximize signal-to-noise of the OSC and RF signal as observed in the spectrum analyzer (SA). The beam displacer generates two probe beams with an offset spacing of 2.8mm and which are orthogonally polarized with respect to each other. The beams travel through the vapor cell centered in between the PPC and are detected by a balanced detector. The output of the balanced detector is split (standard -3dB RF splitter) and sent to an oscilloscope (scope) and SA. The probe signal directed towards the SA and scope is filtered using a band-pass filter (1-20 kHz 3dB passband) to clean-up higher order modes from the coupling of OSC-RF signals. This is most significantly important to clean-up time domain signals, which would otherwise have included higher order components and will distort the appearance of the main tone in the transient waveform of the OSC signal.

The coupler laser driven to $76D_{5/2}$ was frequency locked using a cylindrical cavity (50mm diameter x 100mm length) with finesse of $F_c \sim 15$k. The cavity was placed inside a vacuum housing with multiple internal shields to ensure temperature control. The Pound-Drever-Hall (PDH) technique was used to lock the coupler wavelength via the electronic sideband (ESB) technique and to reduce the linewidth of the coupler laser to <100Hz. The coupler wavelength was $\lambda_c \sim 508.80582$nm, and its detuning is $\Delta_c/2\pi$ = 30MHz when the laser is locked for injected RF measurements in the time domain (probe transmission measurements in time domain). Coupler Rabi frequency and beam diameter was about $\Omega_c$ = 0.85 MHz and 1.3 mm. Probe and coupler laser diodes were a commercial external cavity laser diode (ECDL). The coupler at ~509nm was achieved via frequency-doubling (second harmonic generation).

When EIT frequency spectrum is measured using a scope, frequency sweeping is achieved by piezo electric actuators in the coupler ECDL (extended cavity laser diodes). These are voltage controlled and are driven by a triangular voltage waveform of selected amplitude and frequency. A wavemeter is used to obtain wavelength measurements when coupler frequency is swept. In this case, the coupler linewidth is ~78kHz, the free-running linewidth. Sweep direction is positive (increasing detuning frequency). The scanning rate is set to $2\pi \times 4.88$ MHz/ms and is confirmed with a wavemeter. In swept coupler frequency measurements, OSC is observed qualitatively, and no RF injections are included or measured. When used for time-dependent RF or OSC measurements, the coupler laser is locked to the medium finesse (~15k) cavity, resulting in a linewidth <100Hz as noted previously. In this laser-locked case, OSC and RF (when injected) spectrum is measured. The coupler light is counter-propagated with one of the probe beams. The polarization state of the coupler laser was maintained slightly elliptically polarized using a tunable quarter-wave plate ($\lambda/2$), which is optimized to maximize signal-to-noise of the OSC and RF signal as observed in the SA. The tuning procedure was iterative between the probe and coupler state ($\lambda/4$ tuning), and it was found that improvements in SNR (for OSC and RF) was (<3dB) relative to linearly polarized probe and coupler light in the presented configurations.

When RF signal frequency or amplitude is swept for data collection (passband spectrum in Fig. 1c and 7, and sensitivity measurements in Fig. 8 and 9), sweeping is achieved by programmatic coordination between RF signal source and SA using a computer script.

Additional descriptions of the experimental setup are presented in Supplementary Section 1 that include component/instrument details. The vapor cell design and the study on magnetic field perturbations is given in Supplementary Section 2. The impedance and transfer function of the parallel plate capacitor is given in Supplementary Section 3.

## Data availability

The EIT spectrum as a function of coupler detuning and B-field magnitudes represented in Fig. 2, probe transmission as a function of time for selected B-field values represented in Fig.3, swept frequency analysis and SNR frequency dependence represented in Fig. 7, and the sensitivity data along with curve-fits and errors represented in Fig. 8 are available as source data. All other data are available upon request.

**Figure Legends:**

**Fig. 1: (a)** Experimental protocol, energy diagram, and swept RF frequency spectrum. (a) The probe and coupler laser light are counter-propagated in a Cesium vapor cell to excite atoms to the Rydberg state. A background DC magnetic field between 0.6-5.3 G is generated using a Helmholtz coil (not shown here). A parallel plate capacitor is embedded in the vapor cell, which is used for uniform RF field generation inside the vapor cell. Quarter wave plates are used to fine tune the polarization state of the probe and coupler laser light. A beam displacer is used to develop a reference probe beam for balanced detection to reduce technical noise. **(b)** The probe drives atoms to the first excited state at $6P_{3/2}$ and is detuned by $\Delta_p/2\pi = 22$MHz. The coupler drives the atoms to a principal quantum number of n = 76 at $76D_{5/2}$ and is detuned by $\Delta_c/2\pi = 30$MHz. When the magnetic field is tuned between approximately 4-5G, mode competition is observed between nearby Zeeman sublevels and give rise to limit cycle oscillatory dynamics (OSC). **(c)** The oscillatory dynamics couple the states in a synchronous manner that is observed as a resonance in the probe measured signal at about ~9.8kHz. An injected weak RF signal via the parallel plate capacitor that is close to this resonance modulates the probe signal as the system is sensitized to nearby frequencies. A swept RF frequency shows a finite bandwidth with a highest sensitivity close to the OSC resonance frequency.

**Fig. 2: (a)** EIT spectrum (arbitrary units) as a function of coupler detuning and magnetic flux density with room-temperature Cesium Rydberg vapor. The scanning rate is $2\pi \times 4.88$ MHz/ms. Rabi frequencies of the coupler and probe were held constant. Details of setup is given in Fig. 1 and described in Methods. **(b)** An expanded view of the black box region highlighted in (a) shows that the magnetic flux density magnitude impacts oscillatory magnitude observed by the probe transmission, and that the stability range of the limit cycle oscillations (OSC) in the coupler detuning space.

**Fig. 3:** Probe transmission as a function of time for different magnetic flux density magnitudes. The probe, coupler Rabi frequency was set to $\Omega_{p,c}/2\pi = 13.97$ MHz, 0.85 MHz, the laser wavelengths to drive $6S_{1/2}$-$6P_{3/2}$-$76D_{5/2}$, with $\Delta_{p,c}/2\pi = 22$ MHz, 30 MHz (frequencies locked to observe transients). Between B = 4.3-

4.7G (blue lines), the limit cycle observations (OSC) are stable and sustained. Below B = 4G (red lines), OSC disappears, and background probe laser noise dominates. Above B = 5G (black lines), the OSC is unstable.

**Fig. 4:** The RF modulation of limit cycle oscillations (OSC) as observed by the probe transmission signal. Blue curves are OSC probe transmission with no RF signal, and red curves are a modulated OSC probe transmission due to RF modulation via an electric field. The electric field (red curves) at f = 9.3 kHz and 0.5mV/cm was injected using a parallel-plate capacitor (see Fig. 1 and Methods). Data collection was repeated to show consistency (lighter shaded colors).

**Fig. 5:** Spectral analysis of a driven dissipative time crystal with a strong RF field. **(a)** Both fields (RF and magnetic) are turned on, the limit cycle oscillations are observed, and the strong RF at 11.2kHz is observed. RF and B field magnitudes are ~0.3mV/cm and ~4.5G. **(b)** The RF is turned off and only B-field is on. Here the RF signal is no longer observed in the spectrum. **(c)** The RF is turned on and the B field is off. In (c) and (d), the measured power throughout the band shown is limited by probe laser noise.

**Fig. 6:** An example of progressively weaker RF fields at 11.2kHz. The probe Rabi frequency and magnetic field magnitude is tuned very finely to increase the quality factor of the limit cycle oscillation (OSC). Data collected with a 10Hz resolution bandwidth. Three example curves shown to exemplify the dependence on RF field strength. Curves are shifted vertically (vertical offset) for display.

**Fig. 7:** Swept frequency analysis. **(a)** The electric field transmit frequency is varied over the range of 8-12kHz with a field intensity of ~61.43uV/cm (red curve). This is compared to the case when the field is turned off (blue curve). A passband is observed that drops off rapidly at lower frequencies f<$f_{osc}$, but slowly as f>$f_{osc}$. **(b)** Interpreted signal-to-noise (SNR) ratio showing a peak SNR ~ f<$f_{osc}$ ($f_{osc}$ ~9.8kHz and $f_{peak}$ ~9.5kHz).

**Fig. 8:** Measured probe signal transmission power at the RF frequency for variation in RF field magnitudes. The noise base region (gray rectangular box) is estimated, and two best fit lines are included. The sensitivity is estimated based on data points that are closest but above the noise base region highlighted and based on a 1Hz resolution bandwidth used in the measurement. Two points are selected based on proximity to the best fit lines. These are found to be 1.59uVcm$^{-1}$Hz$^{-1/2}$ and 2.3uVcm$^{-1}$Hz$^{-1/2}$.

## Acknowledgements

The author would like to acknowledge discussions with P. Mao and D. Willey at JPL (Jet Propulsion Laboratory, California Institute of Technology), A. Artusio-Glimpse, N. Prajapati, C. Holloway, and M. Simons at NIST (National Institute of Standards and Technology), and K. Cox, D. Meyer, and P. Kunz at ARL (Army Research Laboratory) as part of the NASA Instrument Incubator Program on Rydberg Radars. The author acknowledges B. Feyissa at JPL (Jet Propulsion Laboratory, California Institute of Technology) for previously developing a script to coordinate triggering of data collection with two laboratory instruments (spectrum analyzer and signal generator). This script was reused in data collection when fine RF E-field sweeps were needed. The research was carried out at the Jet Propulsion Laboratory, California Institute of Technology, under a contract with the National Aeronautics and Space Administration (80NM0018D0004), through

## Author contributions

D.A conceived of the experiment and study reported, configured the atomic systems to include lasers and locking systems, and collected and processed all data reported in the text and figures. D.A. also developed all modeling, theoretical derivations and numerical simulations used or reported.

## Additional information

The author declares no competing interest.

**Supplementary information**

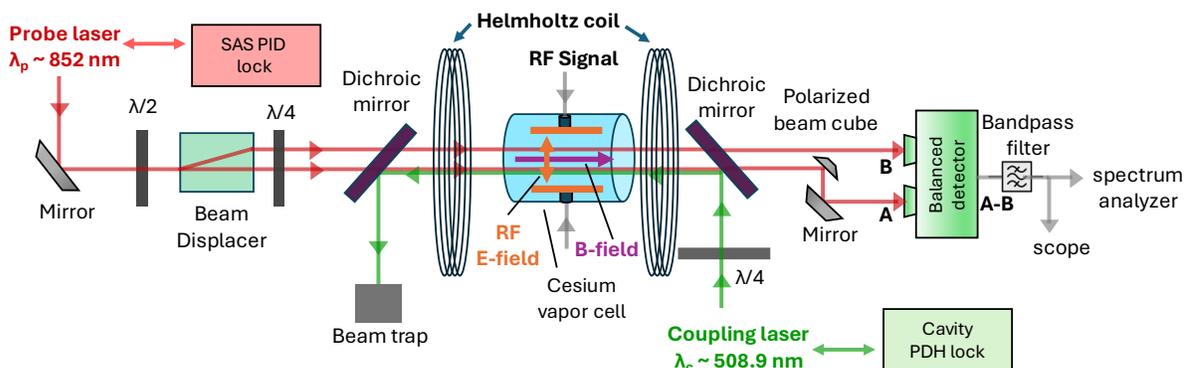

**Extended Data Fig. 1:** A detailed schematic of the experimental setup. The probe (~852 nm) laser (DL Pro Toptica) is frequency locked using a standard saturation absorption spectroscopy (SAS) lock (COSY). The probe laser light drives the $6S_{1/2}$ F=4 to $6P_{3/2}$ F = 4. A half-wave plate ($\lambda/2$) is used to set the polarization state of the light prior toa Calcite beam displacer. The beam displacer generates two probe beams with an offset spacing of 2.8mm and which are orthogonally polarized with respect to each other. The beams travel through the vapor cell and are detected by a balanced detector (Thorlabs PDB250A2). The output of the detector is split and sent to a scope and spectrum analyzer (SA). The RF signal directed towards the SA is filtered using a band-pass filter (1-20 kHz 3dB passband). The coupling light light (~508nm) is generated by a second harmonic generation (SHG) of a ~1016nm light source (TA-MSHG PRO Toptica). The coupler laser light drives the $6P_{3/2}$ F = 4 to $76D_{5/2}$. Detuning's for the probe and coupler is given in the text. Frequency sweeping is achieved by piezo electric actuators in the ECDL (extended cavity laser diodes). These are voltage controlled and are driven by a triangular waveform of selected amplitude and frequency. A wavemeter (HF-ANGSTROM WS/7) is used to obtain wavelength measurements when swept. When used for time-dependent RF measurements, the coupler laser is locked to a medium finesse (~15k) cavity (SLS). The coupler light is counter-propagated with one of the probe beams. A Helmholtz coils that is 30cm is diameter is used to generate uniform B fields with a separation distance of 15cm between the two ends of the coil system. A vapor cell is constructed that has an embedded parallel plate capacitor to generate uniform RF E-fields.

## Supplementary Section 1

### Details of experimental setup and components

The experimental setup (see Fig. 1 and Extended Data Fig. 1) consists of a frequency-locked probe laser (~852 nm, Toptica DL Pro) driving the $6S_{1/2}$ F=4 to $6P_{3/2}$ F=4 transition, with polarization set by a half-wave plate and split into two orthogonally polarized probe beams (2.8 mm apart) using a Calcite beam displacer. The split beam which is linearly polarized is converted to a slightly elliptical polarized beam using a quarter-wave-plate (QWP, see Extended Data Fig. 1). This polarization state is tuned for SNR as noted in the main article text. These beams pass through a vapor cell and are detected by a balanced detector (Thorlabs PDB250A2). The RF signal, filtered (1–50 kHz), is sent to a spectrum analyzer (SA). A coupling laser (~508 nm), generated via SHG of ~1016 nm light (Toptica TA-MSHG PRO), drives the $6P_{3/2}$ F=4 to $76D_{5/2}$ transition. Frequency sweeping is achieved using piezo-controlled ECDLs driven by a triangular waveform. A wavemeter (HF-ANGSTROM WS/7) monitors wavelength, while time-dependent RF measurements use a cavity-locked coupling laser counter-propagated with one probe beam. The cavity was a custom design by SLS Cavity. The cavity was placed inside a vacuum housing with multiple internal shields to ensure temperature control at the cavity of

<1mK/day, with a thermal time constant of about 38 hours. Uniform B-fields are generated by 30 cm Helmholtz coils (by 3B Scientific), and a vapor cell with an embedded parallel plate capacitor (PPC) produces RF E-fields. The PPC and vapor cell was constructed by Precision Glass Blowing. Scope used was a Keysight MSOX3104G. Signal generator used was Agilent E4425B ESG-AP Series Analog RF Signal Generator. Spectrum analyzer used was a N9032B PXA Signal Analyzer.

## Supplementary Section 2

### Vapor cell design and magnetic field perturbations

The vapor cell used has a total length of 56.6 mm and a diameter of 35 mm, with flat windows that are 3.3 mm thick. Parallel plates measure 18 mm × 45 mm and are spaced 15 mm apart. The cell is made of Pyrex, and no buffer gas is utilized. Electrodes are stainless, and a 25mm long cold finger is used. Extended Data Fig. 2 gives the detail drawing for the vapor cell used.

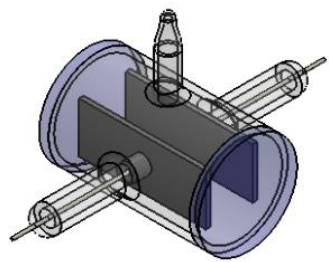

**Extended Data Fig. 2:** Vapor cell geometry. Total length of cell is 56.6mm and diameter of 35mm. Windows are flat and 3.3mm thick. Parallel plates are 18mm x 45mm, and distance between plates are 15mm. Material is Pyrex. No buffer gas is used.

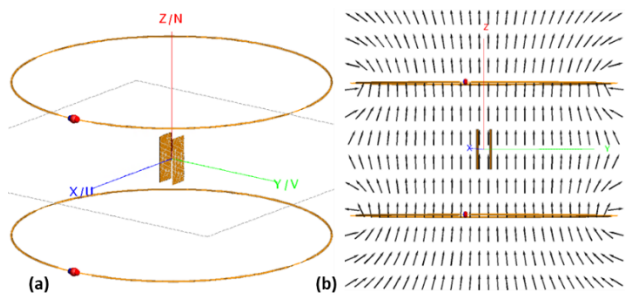

**Extended Data Fig. 3:** Numerical electromagnetic simulations of the Helmholtz geometry and the parallel plate capacitor using Method of Moments (MoM) to study magnetic field perturbations. (a) The simulation setup with the 30cm diameter coils spaced at 15cm. Coils used in the experiment have 20 turns. The parallel plate capacitor is located at the center. (b) The 2D Y-Z cut-plane of magnetic field and field vectors are shown. The field spatial distribution is uniform near and inside the parallel plate capacitor.

Numerical electromagnetic simulations using the Method of Moments (MoM) are used to analyze magnetic field perturbations of the fields by a Helmholtz coil geometry with a central (PPC) parallel plate capacitor (see Extended Data Fig. 3). The field is largely unperturbed. Extended Data Fig. 4 studies the cross-sectional B-field distribution with and without the PPC and we find field percentage variation is ~0.4%. To simplify the simulation, single turn coils were used as opposed to the multi-turn Helmholtz coil design in the experiment, however, this does should not impact the result noted here as the field measurement locations are geometrically far from the wire geometry.

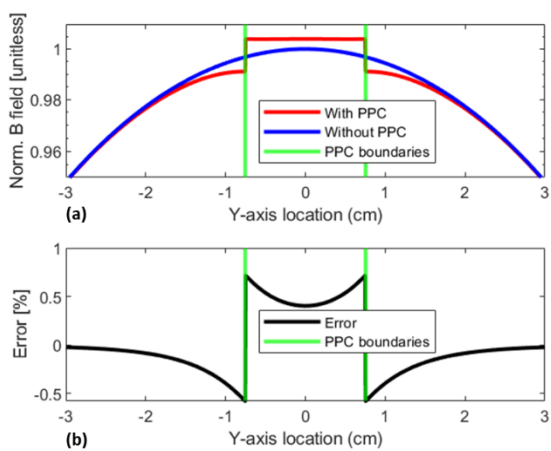

**Extended Data Fig. 4:** Numerical electromagnetic simulations of the magnetic field perturbations along the Y-axis, at X=0. (a) The field is increased inside the parallel plate capacitor (PPC), reduced immediately outside the PPC. (b) The field percentage error inside, relative to the value without the PPC is ~0.4% at approximately Y=0.

**Supplementary Section 3**

**Impedance, transfer function for the parallel plate capacitor, and conversion to field generated**

A sensitive RF impedance analyzer (Keysight E4990A Impedance Analyzer) is used to measure the DC to 100 MHz impedance of the PPC embedded inside the vapor cell. Extended Data Fig. 5 gives the result of this frequency dependent impedance measurement. A theoretical circuit model is fitted to this data at the low-frequency regime (<10MHz, dashed lines in Extended Data Fig. 5a). This model is based on a parallel resistor and capacitor model, where the capacitance is the standard $C = \varepsilon_r \varepsilon_0 A/d$, where A is the area of the PPC plates, d the distance between the two PPC plates, and $\varepsilon_r$ the relative dielectric constant. Using the low-frequency and conductive limit with dielectric relaxation dominating, $\varepsilon_r \varepsilon_0 \sim \sigma \tau$, where $\sigma$ is the conductivity and $\tau$ the relaxation time constant, and $\sigma$ is estimated via DC measured resistance, and $\tau \sim 8$ns estimated for best curve fit in Extended Data Fig. 5a. RMS error of the fit is ~3.6% up to 10 MHz. Weak resonances are seen in the 50-80MHz regime likely due to inductances in the transmission line due to feeding structures.

The magnitude of impedance of the load ($|Z|$) is shown in in Extended Data Fig. 6a up to 10 MHz. The absolute value of the voltage transfer function from a 50

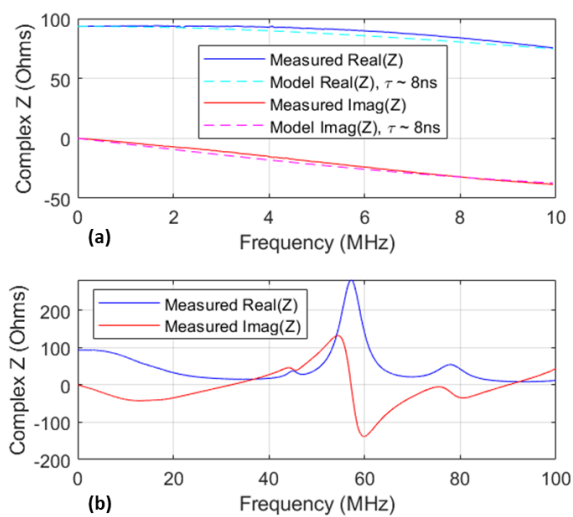

**Extended Data Fig. 5:** (a) Measured impedance up to 10MHz and fit with expected model, showing a best fit at a relaxation time constant of $\tau \sim 8$ns. The model is in good agreement with the impedance data (RMS error ~3.6%). (b) The impedance over a larger range up to 120MHz is given. Weak resonances above 40MHz are attributed to higher order inductances in the feed structure.

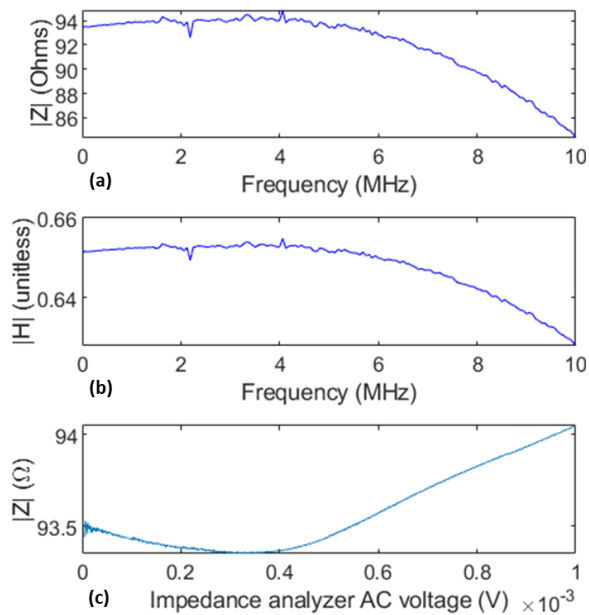

**Extended Data Fig. 6:** (a) Magnitude of impedance (|Z|) up to 10MHz. (b) The absolute value of the voltage transfer function from a 50 Ohm source to the parallel plate capacitor up to 10MHz. (c) Magnitude of impedance as a function of AC voltage applied by the impedance analyzer at 10kHz, showing a change of about ~0.6 Ohms between 1mV and 100uV.

Ohm RF source ($R_s$) to the PPC is calculated using a standard voltage divider (by $H = Z/(Z + R_s)$) and presented in Extended Data Fig. 6b (|H| ~0.62, unitless). The E-field generated is then calculated as $|H| V_{RF}/d\sqrt{2}$, where $V_{RF}$ is the voltage applied by the signal generator, and $d$ is the spacing between the plates on the PPC. Additionally, the magnitude of impedance is plotted as a function of the AC voltage applied by the impedance analyzer at 10 kHz, revealing a change of approximately 0.6 Ohms between 1 mV and 100 μV.

**Supplementary Section 4**
**Error curves from best fit models**

The peak power at $f_{RF}$ =9.5kHz is measured in the sensitivity measurements presented in Fig. 8 within the main article text. The OSC frequency for this measurement is identical to that used in Fig. 7 of the main article text. The shaded grey region represents the estimated noise floor, primarily influenced by probe laser noise. This is estimated by the mean value of the weakest 7 measurement points, which correspond to field values between 0.307–0.921 μV/cm. The grey region (noise-base region) in Fig. 8 within the main article text has a y-axis extend limited programmatically by the mean value of the read-out peak power in the SA between -109.97dB and -113.22dB. Measurements were conducted four times over two days, with error bars indicating the standard error of the mean. The standard error of the mean (SEM) quantifies the precision of the sample mean as an estimate of the true population mean, calculated by dividing the sample standard deviation by the square root of the number of repeated measurements. Two curve fits to the data are shown in Fig. 8 of the main article text: a log-log fit (grey line) and an offset log-log fit (red curve). The log-log fit uses an equation $P = b_1 \log_{10} x + b_2$, whereas the offset log-log fit uses an equation $P = b_1 \log_{10}(x + b_3) + b_2$, where $b_n$ constants are solve in independently in both cases via a non-linear least square optimization using the trust-region-reflective algorithm and x is the field magnitude. The offset log-log model is introduced as it produces an improved (lower error) fit to the mean valued data points. In model fitting, the noise corrupted data values are not used and only data from 1.597-61.4 μV/cm is used to solve for the $b_n$ constants. The estimated errors for the best-fit curves at each data point are shown in Extended Data Fig. 7, with the log-log fit showing an RMS error of approximately 2dB and the offset log-log fit exhibiting an error of 0.63dB.

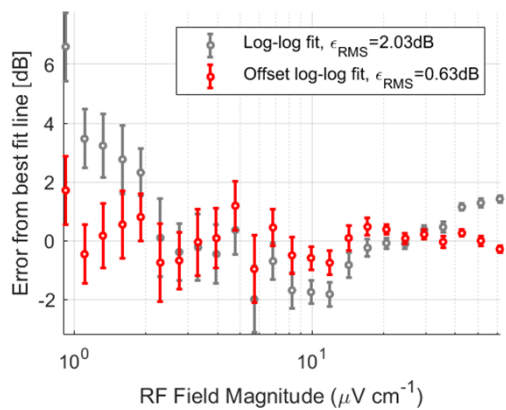

**Extended Data Fig. 7:** Estimated error in best fit curves shown in Fig. 8. The RMS error is given in the legend, where the log-log fit gives an RMS error of ~2dB, and the offset log-log fit gives an error of 0.63dB.